\documentclass[journal,twoside,web]{ieeecolor}
\usepackage{lcsys}
\usepackage{cite}
\usepackage{amsmath,amssymb,amsfonts,amstext}
\usepackage{graphicx}
\usepackage{textcomp}
  \def\BibTeX{{\rm B\kern-.05em{\sc i\kern-.025em b}\kern-.08em
     T\kern-.1667em\lower.7ex\hbox{E}\kern-.125emX}}
\usepackage{xcolor}
\usepackage[utf8]{inputenc}
\usepackage{babel}
\usepackage{balance}
\usepackage{comment}
\usepackage[position=bottom]{subfig}

\usepackage{mathtools}
\usepackage{float}

\usepackage{amsthm}
\newtheorem{theorem}{Theorem}
\newtheorem{lemma}{Lemma}

\newtheorem{problem}{Problem}

\usepackage{algorithm}
\usepackage{algorithmicx} 
\usepackage{algpseudocode} %

\usepackage{todonotes}

\makeatletter
\let\NAT@parse\undefined
\makeatother

\title{Combined Optimal Routing and Coordination of Connected and Automated Vehicles}

\author{Heeseung Bang, \textit{Student Member, IEEE}, Behdad Chalaki, \textit{Student Member, IEEE},\\Andreas A. Malikopoulos, \textit{Senior Member, IEEE}
    \thanks{This research was supported by ARPAE's NEXTCAR program under the award number DE-AR0000796.}
	\thanks{The authors are with the Department of Mechanical Engineering, University of Delaware, Newark, DE 19716, USA. (emails: \tt\small{heeseung@udel.edu}; \tt\small{bchalaki@udel.edu}; \tt\small{andreas@udel.edu}.)}
}

\date{February 2022}

\begin{document}

\maketitle
\thispagestyle{empty}
\begin{abstract}

In this letter, we consider a transportation network with a 100\% penetration rate of connected and automated vehicles (CAVs) and present an optimal routing approach that takes into account the efficiency achieved in the network by coordinating the CAVs at specific traffic scenarios, e.g., intersections, merging roadways, and roundabouts.
To derive the optimal route of a travel request, we use the information of the CAVs that have already received a routing solution. This enables each CAV to consider the traffic conditions on the roads.
The solution of any new travel request determines the optimal travel time at each traffic scenario while satisfying all state, control, and safety constraints.
We validate the performance of our framework through numerical simulations. To the best of our knowledge, this is the first attempt to consider the coordination of CAVs in a routing problem.
\end{abstract}


\section{Introduction}
\IEEEPARstart{D}{ue} to the increasing population and travel demand, traffic congestion has dramatically increased over the last decade \cite{urbanization2018,Schrank2019}.
As one of the promising ways to alleviate traffic congestion, connected and automated vehicles (CAVs) have received great attention. 
The majority of current research efforts have considered coordination and control of CAVs to improve fuel efficiency and alleviate traffic congestion. 
Some  approaches provide coordination of CAVs at different traffic scenarios such as signal-free intersections \cite{Dresner2008,chalaki2020TITS,Au2015,Kumaravel:2021uk}, merging roadways \cite{Athans1969,Papageorgiou2002,Ntousakis:2016aa}, and corridors \cite{Zhao2018ITSC,Lee2013,mahbub2020decentralized}. 
More recent efforts have focused on developing learning techniques for coordinating CAVs \cite{chalaki2020ICCA,davarynejad2011motorway,nassef2020building,wang2019q}. 

From a macroscopic perspective, several papers have addressed the routing problem focusing on different types of vehicles, e.g., hybrid electric vehicles (HEVs), plug-in HEVs \cite{huang2020eco,salazar2019optimal}. However, in computing the optimal energy management strategy for the vehicles, these approaches neglect traffic congestion, or the impact of the other vehicles on their routing solution.
To solve the dynamic vehicle-assignment problem, Chen and Cassandras \cite{chen2020optimal} developed an event-driven receding horizon control scheme in a mobility-on-demand system.
The solution minimizes customers' waiting time and travel time but does not consider traffic congestion on the road.
Tsao et al. \cite{tsao2019model} used model predictive control to find the optimal routes for ride-sharing by considering a constant value for the travel time of each road segment.

There have been other efforts on the routing problem considering traffic flow and congestion on the roads.
Smith et al. \cite{smith2013rebalancing} presented a rebalancing strategy for mobility-on-demand system vehicles and drivers. Rather than focusing on each vehicle model, they used a fluid-approximated model to capture the vehicle flow.
Salazar et al. \cite{salazar2019congestion} developed a method for congestion-aware routing and rebalancing. They considered the flow of the vehicles and adopted the volume-delay function to model traffic congestion.
Wollenstein-Betech et al. \cite{wollenstein2020congestion} used the same approach but with a more accurate approximation of the volume-delay function and applied it to a mixed traffic scenario. The authors formulated a game-theoretic problem between CAVs and privately-owned vehicles and considered that the latter solely seek to minimize their costs. 
Other research efforts extended the work in \cite{salazar2019congestion,wollenstein2020congestion} and applied it to a more complex network  \cite{wollenstein2021routing} and to situations where charging scheduling of electric CAVs is required \cite{bang2021AEMoD}.
In these papers, the routing problems were solved with respect to the vehicle flow and neglected the actual movement of each CAV at the microscopic level.
In addition, they assumed the travel demand to be time-invariant so that the traffic conditions would not change. Therefore, the estimation of traffic congestion presented in \cite{salazar2019congestion,wollenstein2020congestion,wollenstein2021routing,bang2021AEMoD} cannot reflect real traffic conditions caused by each CAV's movement.

\begin{figure}[!h]
    \centering
    \includegraphics[width=1\linewidth]{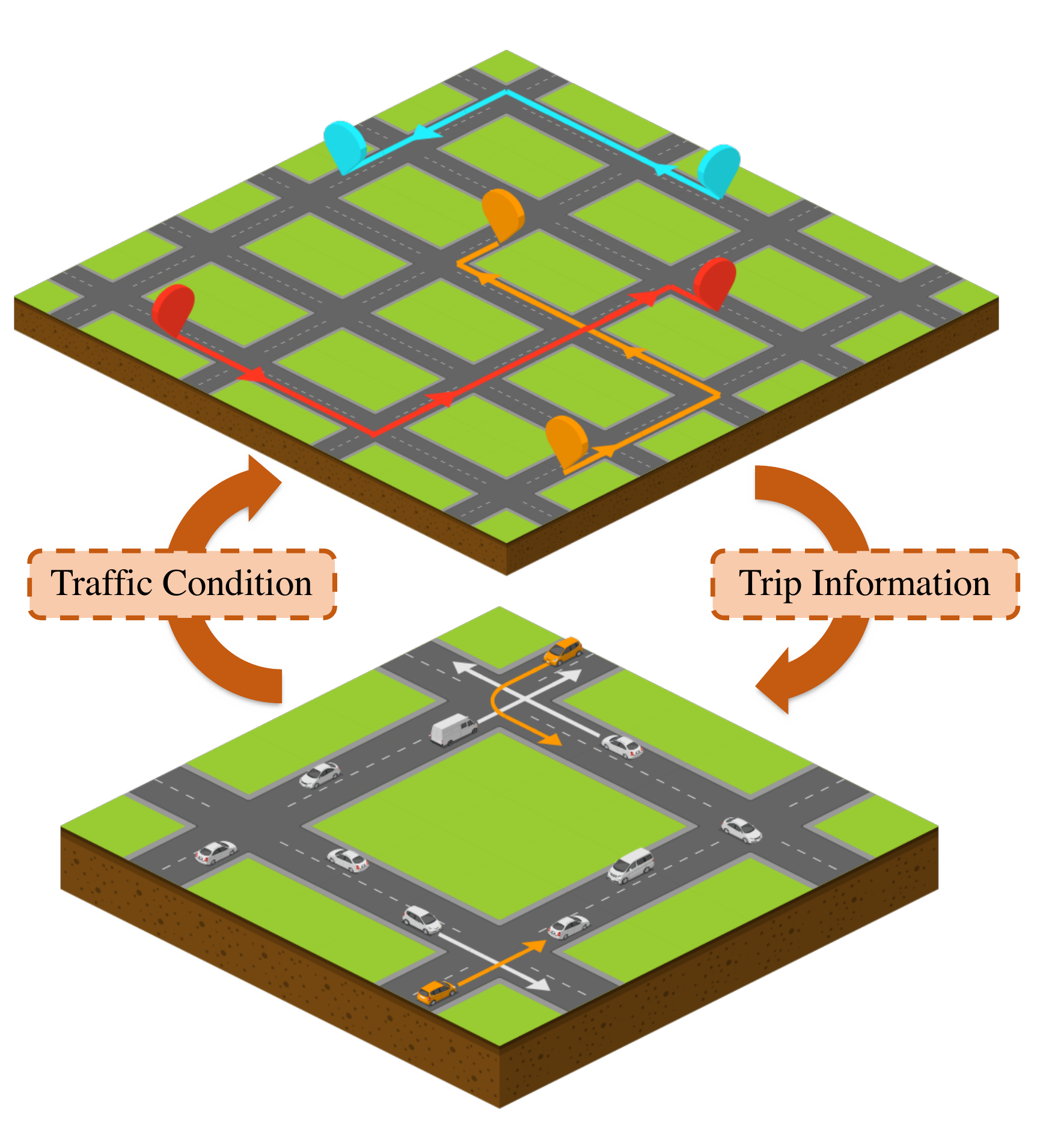}
    \caption{Combined framework of routing and coordination of CAVs.}
    \label{fig:combined}
\end{figure}

In this letter, we propose a framework that combines routing and coordination of CAVs (Fig.~\ref{fig:combined}).
Although there has been a significant amount of efforts on both routing and coordination of CAVs, none of these efforts has considered the associated challenges simultaneously.
While most of the existing papers neglect (or just estimate) traffic congestion for the routing problem, the proposed framework allows us to receive the actual traffic conditions of the entire network at all time (including future time) by using the coordinated trajectories of CAVs.

We consider a $100$\% penetration rate of CAVs on the roads and solve the routing problem in the order of travel requests.
To derive the optimal route of a travel request, we use the information of the  CAVs that have already received a routing solution. This enables each CAV to consider the traffic conditions on the roads.
Given the trajectories of CAVs resulting from the routing solutions, the solution of any new travel request determines the optimal travel time at each traffic scenario while satisfying all state, control, and safety constraints.

The main contributions of this letter are (1) bridging the gap between macroscopic and microscopic traffic analysis, (2) establishing a framework that captures traffic congestion in the routing problem, and (3) demonstration of how the combined routing and coordination of CAVs can improve performance in the road network.
To the best of our knowledge, considering the actual trajectory of each vehicle in the routing problem has not yet been reported in the literature to date.

The remainder of this letter is organized as follows.
In Section \ref{sec:modeling}, we introduce the system model from the network to CAVs.
In Section \ref{sec:routing}, we formulate the optimal routing problem at a network level and provide a solution approach, while in Section \ref{sec:coordination}, we present a coordination framework of CAVs at a signal-free intersection.
We validate the effectiveness of our method through simulations in Section \ref{sec:simulation}. We draw concluding remarks and discuss potential directions for future research in Section \ref{sec:conclusion}.


\section{System Modeling}   \label{sec:modeling}

\begin{figure}[t]
    \centering
    \includegraphics[width=\linewidth]{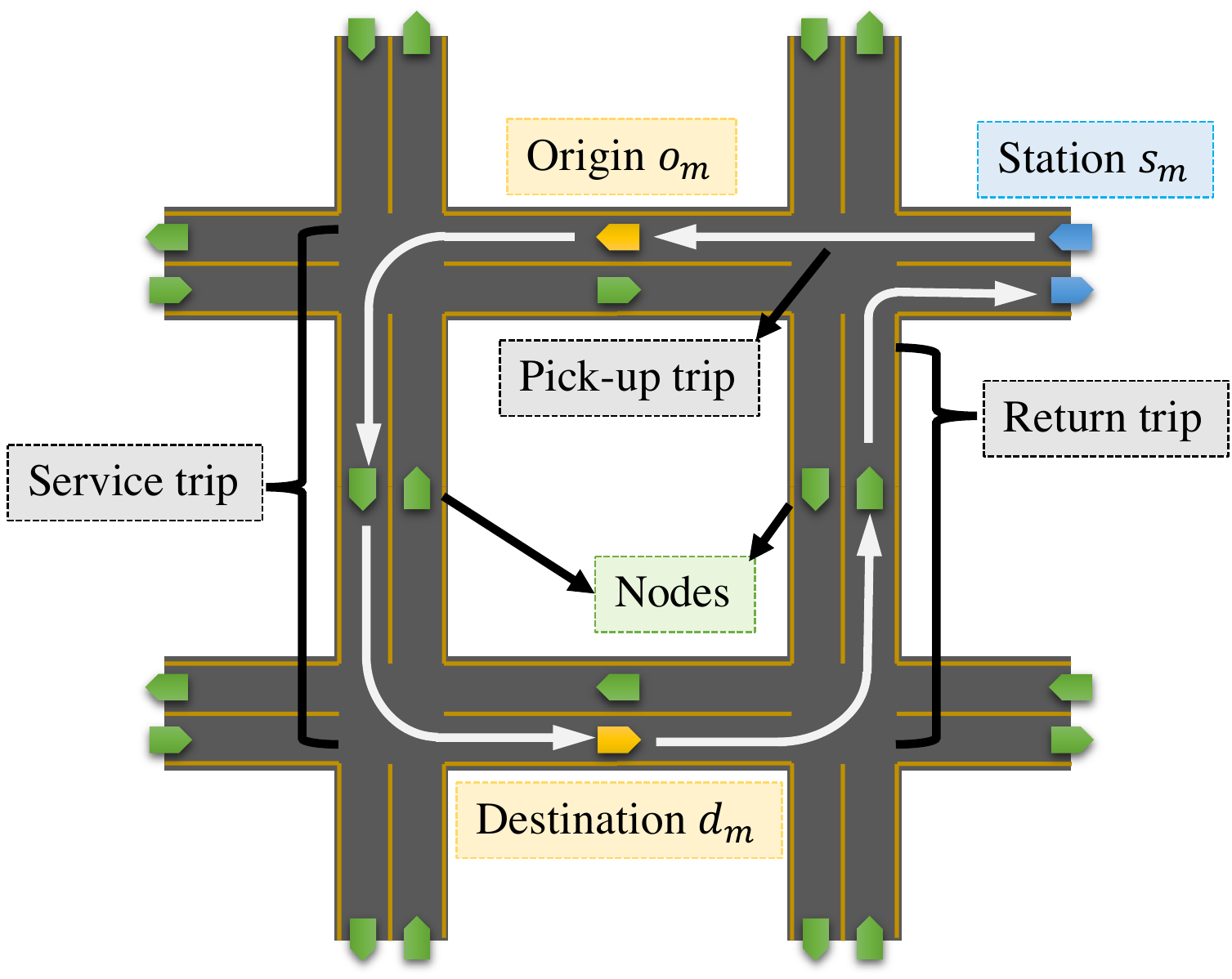}
    \caption{Part of the road network. Polygons illustrate nodes and white arrows represent some of the directed edges. The whole trip for one travel request consists of pick-up trip, service trip, and return trip.}
    \label{fig:network}
\end{figure}

\subsection{Road Network and Travel Request}

Although the proposed framework can be applied in a transportation network with any traffic scenario, e.g., crossing intersections, merging at roadways and roundabouts, cruising in congested traffic, and passing through speed reduction zones, we restrict our attention to networks with intersection and single-lane roads to simplify the exposition and notation.
Thus, we consider a road network that consists of intersections and single-lane roads (Fig. \ref{fig:network}). 
Each intersection includes a \textit{coordinator} that stores information about the intersection's geometry and CAVs' trajectories. The coordinator only acts as a database for the CAVs crossing it and does not make any decision.
We model the road network with a directed graph $\mathcal{G} = (\mathcal{V},\mathcal{E})$, where $\mathcal{V}$ denotes a set of nodes and $\mathcal{E}\subset\mathcal{V}\times\mathcal{V}$ denotes the set of roads. 
The network includes a \textit{routing decision unit} (RDU) that receives all the travel information and finds the optimal route for each travel request.
The RDU communicates with coordinators to get exact traffic conditions at the intersections.

We define a set of station nodes $\mathcal{S}\subset\mathcal{V}$, where all the CAVs are initially located.
Once a travel request has been made, a single CAV from a particular station $s\in\mathcal{S}$ must be assigned to the travel request.
Although the travel-vehicle assignment is another interesting problem to explore \cite{li2020trip,mori2021request}, in this letter, we consider the assignment to be given from a higher-level decision layer and assume that CAVs always return to the same station where they are dispatched from.

The RDU receives information on travel requests whenever a customer calls for a ride.
Let $\mathcal{M}(t)\subset\mathbb{N}$ be the set of all travel requests and $\mathcal{N}(t)\subset\mathbb{N}$ be the set of CAVs assigned to each travel request at time $t\in\mathbb{R}_{\geq0}$.
For each travel request $m\in\mathcal{M}(t)$, we have an information tuple $\mathbb{I}(m) = (s_m,o_m,d_m,t_{s_m})$, which consists of the assigned station $s_m\in\mathcal{S}$, origin $o_m$, destination $d_m$, and the initial time $t_{s_m}$ at the station $s_m$.
The RDU finds the time-optimal routes in the order of the travel requests, and it searches for three different trips (Fig. \ref{fig:network}) for each travel request $m\in\mathcal{M}(t)$, namely, a \textit{pick-up trip} $(s_m\to o_m)$, a \textit{service trip} $(o_m\to d_m)$, and a \textit{return trip} $(d_m \to s_m)$.
For each trip, we let $n^\mathrm{i}$ and $n^\mathrm{f}$ denote initial and final node, respectively, e.g., $n^\mathrm{i} = s_m$ and $n^\mathrm{f} = o_m$ for a pick-up trip.

\subsection{Modeling of Connected and Automated Vehicles at Signal-Free Intersections}

\begin{figure}[t]
    \centering
    \includegraphics[width=1\linewidth]{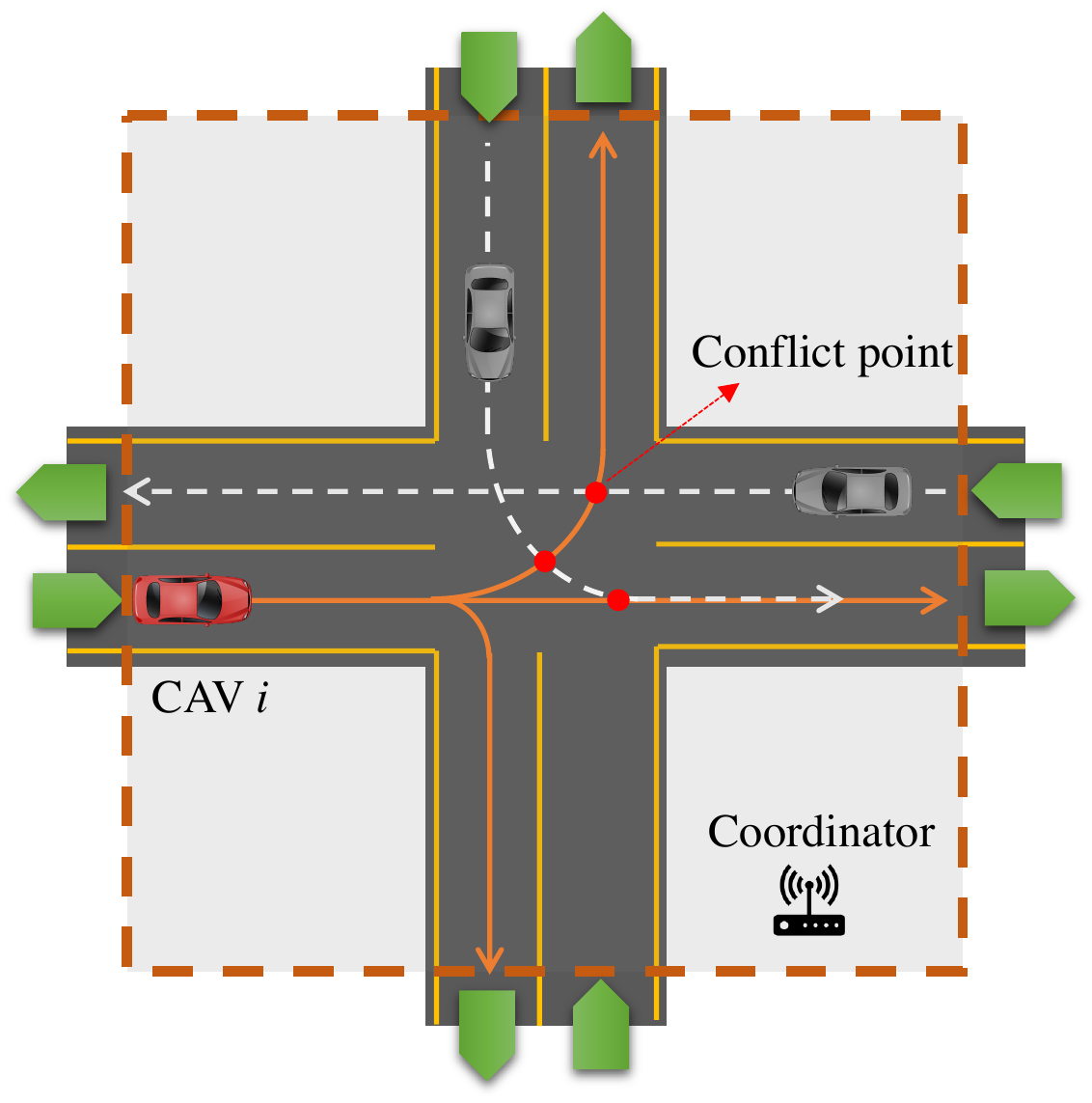}
    \caption{Coordination of CAV $i$ at intersection considering other CAVs' trajectories. Solid lines show all the three possible paths for CAV $i$.}
    \label{fig:intersection}
\end{figure}

We consider a signal-free intersection (Fig. \ref{fig:intersection}). The four-way intersection consists of four inlet nodes and four outlet nodes. Neglecting U-turns at each intersection, we have twelve unique paths for these inlets and outlets.
Once a CAV reaches the inlet node of an intersection, the CAV is considered to belong to that intersection.
Upon exit of the outlet node, the CAV is considered to belong to the adjacent traffic scenario, e.g., another intersection or just a road.
There are \textit{conflict points} in the intersection (Fig. \ref{fig:intersection}), where two paths intersect at which the CAVs may have a lateral collision.
We define the index set of conflict points $\mathcal{C}\subset\mathbb{N}$.

We model each CAV $i\in\mathcal{N}(t)$ as a double integrator
\begin{equation}
    \begin{aligned}
    \dot{p}_i(t) &= v_i(t),\\
    \dot{v}_i(t) &= u_i(t),
    \end{aligned}
\end{equation}
where $p_i(t)\in\mathbb{R}$, $v_i(t)\in\mathbb{R}$, and $u_i(t)\in\mathbb{R}$ denote position, speed, and control input at time $t$, respectively.
In our modeling framework, we impose the following constraints for each CAV $i\in\mathcal{N}(t)$
\begin{align}
    u_{i,\text{min}} & \leq u_i(t) \leq u_{i,\text{max}}, \label{eqn:ulim}\\
    0 < v_{\text{min}} & \leq v_i(t) \leq v_{\text{max}}, \label{eqn:vlim}
\end{align}
where $u_{i,\text{min}}$, $u_{i,\text{max}}$ are the minimum and maximum control inputs and $v_{\text{min}}$, $v_{\text{max}}$ are the minimum and maximum allowable speed, respectively.

The coordinator at the intersection monitors the trajectory information of all CAVs crossing the intersection and broadcasts the traffic conditions back to the RDU. Thus, the RDU can derive the optimal route for each travel request by considering the traffic conditions at each intersection.
Let $\mathcal{R}$ denote the set of intersections and $\mathcal{N}_r(t)\subset\mathcal{N}(t)$ be a set of CAVs traveling through the intersection $r\in\mathcal{R}$.
For each CAV $i\in\mathcal{N}_r(t)$, let $t_i^0$ and $t_i^f$ be the time that CAV $i$ enters and exits the intersection, respectively. 
Since the RDU derives the optimal routes in the order of travel requests, by the time CAV $i$ accommodates a travel request, the trajectories of CAVs $\{1,\dots,i-1\}$ have been already determined and shared with the coordinator.

\section{Optimal Routing Scheme} \label{sec:routing}

In this section, we present the routing problem and solution approach.
Since routes are determined in the order of the requests, we focus on a routing of CAV $i\in\mathcal{N}(t)$ and the routes and trajectory information of CAVs $1$ to $i-1$.
We first formulate the routing problem using a dynamic programming (DP) decomposition to derive the shortest-time route.
Let the state of the system at stage $l=0,\dots,N$, $N\in\mathbb{N},$ consist of the (1) node $x_l\in\mathcal{V}$ and (2) arrival time $t_{x_l}\in\mathbb{R}_{\geq0}$ at that node.
The control $y_l\in\mathcal{Y}(x_l)$ at stage $l$ is to select a node that is connected to the current node $x_l$, i.e., $\mathcal{Y}(x_l)=\{v\in\mathcal{V}~\big|~(x_l,v)\in\mathcal{E}\}$.
The evolution of the state of the system is given by
\begin{align}
    x_{l+1} &= y_l,~~l=0,\dots,N-1,\\
    t_{x_{l+1}} &= t_{x_l}+g_l(x_l,y_l,t_{x_l}),~~l=0,\dots,N-1,
\end{align}
where $g_l(\cdot)$ is the travel time from $x_l$ to $y_l$ at time $t_{x_l}$.

\begin{problem}[DP for routing] \label{prb:DP}
    For each trip, the shortest-time route is obtained by solving the following iterative equation
    \begin{equation}
        J_l(x_l,t_{x_l}) = \min_{y_l\in\mathcal{Y}(x_l)} \big[g_l(x_l,y_l,t_{x_l})+J_{l+1}(x_{l+1},t_{x_{l+1}})\big],   \label{eqn:dp}
    \end{equation}
    for $l=0,\dots,N-1$, where the terminal cost is given by $J_N(x_N,t_{x_N})=g_N(x_N,t_{x_N})=0$.
\end{problem}

Solving Problem \ref{prb:DP} is computationally expensive because it requires to solve the problem for all possible terminal time $t_{x_N}$ and find the optimal cost $J^*(n^\mathrm{f},t_{n^\mathrm{f}}^*)$.
Besides, the continuous state $t_{x_l}$ requires discretization, which dramatically increases the number of possible states at each backward step.
To reduce the computational cost, we convert Problem \ref{prb:DP} to an equivalent forward dynamic programming (FDP) problem with less number of possible states.

\begin{lemma}   \label{lem:FDP}
    For a deterministic finite-state DP problem, a backward DP algorithm can be converted into an FDP algorithm.
\end{lemma}

\begin{proof}
    See p.70 in \cite{bertsekas2012dynamic}. 
\end{proof}

The FDP uses the following iterative equation
\begin{equation}
    \Tilde{J}_{l+1}(x_{l+1},t_{x_{l+1}}) = \min_{x_l\in\mathcal{X}(y_l)} \big[g_l(x_l,y_l,t_{x_l})+\Tilde{J}_{l}(x_{l},t_{x_{l}})\big], \label{eqn:fdp}
\end{equation}

for $l=0,\dots,N-1$, where $x_{l+1}=y_l$, $t_{x_{l+1}}=g_l(x_l,y_l,t_{x_l})+t_{x_l}$, and $\mathcal{X}(y_l)=\{v\in\mathcal{V}~\big|~(v,y_l)\in\mathcal{E}\}$.
The initial cost is given by $\Tilde{J}_0(x_0,t_{x_0})=0$, where $x_0=n^\mathrm{i}$ and $t_{x_0}=t_{n^\mathrm{i}}$.
In this case, the cost function represents the shortest time-to-arrive at the current state.

\begin{lemma}   \label{lem:time}
    For the routing problem of CAVs traveling in a single-lane road network, the following inequality always holds: $g_l(x_l,y_l,t_{x_l})+\Tilde{J}_{l}(x_{l},t_{x_{l}}) < g_l(x_l,y_l,t_{x_{l}}^\prime)+\Tilde{J}_{l}(x_{l},t_{x_{l}}^\prime)$ for any $l=0,\dots,N-1$, $(x_l,y_l)\in\mathcal{E}$, if $t_{x_{l}} < t_{x_{l}}^\prime$.
\end{lemma}

\begin{proof}
    The inequality directly follows the physical condition of a single-lane road network where a CAV cannot overtake the preceding CAVs on the same road.
\end{proof}

From \eqref{eqn:fdp}, we can conclude that $\Tilde{J}_{l+1}(x_{l+1},t_{x_{l+1}}) = t_{x_{l+1}}$.
By Lemma \ref{lem:time}, a solution to the single iteration of \eqref{eqn:fdp} is always obtained from the shortest arrival time at certain node $x_{l}$.
Thus, by considering the shortest travel time, the arrival time can be eliminated from the state without affecting the solution.
Next, we define the revised FDP problem.

\begin{problem}[FDP] \label{prb:DP2}
    The shortest-time route is obtained by solving the following FDP problem
    \begin{equation}
        \Tilde{J}_{l+1}(x_{l+1}) = \min_{x_l\in\mathcal{X}(y_l)} \big[\Tilde{g}_l(x_l,y_l,\Tilde{J}_{l}(x_{l}))+\Tilde{J}_{l}(x_{l})\big],   \label{eqn:dp2}
    \end{equation}
    ~for $l=0,\dots,N-1$, where the initial cost is given by $\Tilde{J}_0(x_0)=t_{n^\mathrm{i}}$ and $\Tilde{g}_l(\cdot)$ is the short travel time.
\end{problem}

\begin{theorem}
    The optimal solution to the Problem \ref{prb:DP2} is also solution to the Problem \ref{prb:DP}.
\end{theorem}
\begin{proof}
The proof follows from Lemmas \ref{lem:FDP} and \ref{lem:time}. 
\end{proof}

Note that the shortest travel time $\Tilde{g}_l(\cdot)$ is computed at the coordination level, which will be discussed in Section IV. For the coordination, each CAV finds $\Tilde{g}_l(\cdot)$ and the route under the condition that its trajectory does not affect the pre-planned trajectories of the other CAVs. This implies that the solution to Problem 2 for a new travel request does not affect the existing solutions of the other CAVs.

\section{Coordination of Connected and Automated Vehicles} \label{sec:coordination}

For the coordination of CAVs at a signal-free intersection $r\in\mathcal{R}$, we employ the framework presented in \cite{chalaki2021Reseq}.
In this framework, for each CAV, we seek to derive an energy-optimal trajectory while minimizing travel time and satisfying state, control, and safety constraints. If such trajectory exists, it is guaranteed to satisfy all the constraints while reducing energy consumption and travel time.

The energy-optimal unconstrained trajectory for each CAV $i\in\mathcal{N}_r(t)$ is \cite{Malikopoulos2020}
\begin{align} \label{eq:optimalTrajectory}
    u_i(t) &= 6 a_i t + 2 b_i, \notag \\
    v_i(t) &= 3 a_i t^2 + 2 b_i t + c_i, \\
    p_i(t) &= a_i t^3 + b_i t^2 + c_i t + d_i, \notag
\end{align}
where $a_i, b_i, c_i$, and $d_i$ are constants of integration, which can be computed using the following boundary conditions
\begin{align}
     p_i(t_i^0) &= 0,\quad  v_i(t_i^0)= v_i^0 , \label{eq:bci}\\
     p_i(t_i^f)&=p_i^f,\quad u_i(t_i^f)=0. \label{eq:bcf}
\end{align}
The final speed $v_i(t_i^f)$ varies with respect to $t_i^f$, hence we have $u_i(t_i^f)=0$ \cite{bryson1975applied}.
For the detailed derivation of the energy-optimal trajectory, see \cite{Malikopoulos2020}.

Next, we consider safety constraints.
To guarantee rear-end safety between CAV $i\in\mathcal{N}_r(t)$ and a preceding CAV $k\in\mathcal{N}_r(t)$ in the same path, we impose the following constraints,
\begin{equation}
    p_k(t)-p_i(t) \geq \delta_i(t) = \rho + \varphi \cdot v_i(t),  \label{eqn:rearend1}
\end{equation}
where $\delta_i(t)$ is the safety distance depending on the speed of CAV $i$, $\rho$ is the standstill distance, and $\varphi$ is a reaction time.
We also consider the rear-end safety between CAV $i$ and a preceding CAV $j$ as
\begin{equation}
    p_i(t)-p_j(t) \geq \delta_j(t) = \rho + \varphi \cdot v_j(t).\label{eqn:rearend2}
\end{equation}
For the lateral collision at a conflict point, we consider two different scenarios. Suppose CAV $k\in\mathcal{N}_r(t)$ is the CAV that already planned its trajectory and passes a conflict point that might cause a lateral collision with CAV $i$.
Then, CAV $i$ can pass the conflict point either before or after CAV $k$. In the first case, the trajectory of CAV $i$ must satisfy
\begin{equation}
    p_i^c - p_i(t) \geq \delta_i(t),~\forall t \in [t_i^0,t_k^c], \label{eqn:lateral1}
\end{equation}
where $p_i^c\in\mathbb{R}$ is the location of the conflict point $c\in\mathcal{C}$ on CAV $i$'s path, and $t_k^c$ is the known time that CAV $k$ reaches at the conflict point $c\in\mathcal{C}$.
In the second case, where CAV $i$ passes the conflict point after CAV $k$, the constraint becomes
\begin{equation}
    p_k^c - p_k(t) \geq \delta_k(t) = \rho+\varphi\cdot v_k(t),~\forall t\in[t_k^0,t_i^c], \label{eqn:lateral2}
\end{equation}
where $p_k^c\in\mathbb{R}$ is the location of the conflict point on CAV $k$'s path, and $t_i^c$ is the time when CAV $i$ reaches the conflict point $c\in\mathcal{C}$, which is determined by the trajectory of CAV $i$.
The position $p_i(t)$ in \eqref{eq:optimalTrajectory} is strictly increasing because of \eqref{eqn:vlim}.
Thus, the inverse $t_i(\cdot)=p_i^{-1}(\cdot)$ always exists.
We call this function the \textit{time trajectory} of CAV $i$ \cite{Malikopoulos2020}.
From this function, we obtain the time $t_i^c = p_i^{-1}(p_i^c)$ at which CAV $i$ arrives at the conflict point $c$ along the energy optimal trajectory \eqref{eq:optimalTrajectory}.

To ensure lateral safety, either \eqref{eqn:lateral1} or \eqref{eqn:lateral2} must be satisfied.
Therefore, we impose the lateral safety constraint by taking the minimum of \eqref{eqn:lateral1} and \eqref{eqn:lateral2}, i.e.,
\begin{align}\label{eq:lateralMinSafety}
    \min \Bigg\{ &\max_{t\in[t_i^0, t_k^c]} \{ \delta_i(t) + p_i(t) - p_i^c\}, \notag\\
            &\max_{t\in[t_k^0, t_i^c]} \{ \delta_k(t) + p_k(t) - p_k^c \}   \Bigg\} \leq 0. 
\end{align}

To find the minimum exit time $t_i^f$, we define the feasible set $\mathcal{T}_i=\left[\underline{t}_i^{f}, \overline{t}_i^f\right]$, where $\underline{t}_i^{f}$ is earliest exit time and $\overline{t}_i^f$ is latest exit time that CAV $i$ can exits the intersection with an unconstrained energy optimal trajectory \eqref{eq:optimalTrajectory}. This set can be constructed using speed and control input limits \eqref{eqn:ulim}, \eqref{eqn:vlim} and boundary conditions \eqref{eq:bci}, \eqref{eq:bcf} \cite{chalaki2020experimental}.

\begin{problem}
To find minimum exit time, each CAV $i\in\mathcal{N}_r(t)$ solves the following optimization problem
    \begin{align}\label{eq:tif}
        &\min_{t_i^f\in \mathcal{T}_i} t_i^f \\
        \emph{subject to: }&
         \eqref{eq:optimalTrajectory}-  \eqref{eqn:rearend2}, \eqref{eq:lateralMinSafety} \notag.
    \end{align}
    \label{pb:timeMinRP}
\end{problem}
The solution to the Problem \ref{pb:timeMinRP} can be directly used for $\Tilde{g}_l(\cdot)$ in \eqref{eqn:dp2}.
CAV $i$ solves Problem \ref{pb:timeMinRP} using Algorithm \ref{Alg:Coordination}.
Algorithm \ref{Alg:Coordination}, returns $-1$ whenever $t_i^f$ exceeds the maximum allowable exit time with the unconstrained energy optimal trajectory.
This implies that there is no unconstrained energy optimal trajectory that satisfies all state, control, and safety constraints.
We can assign a big number $B\in\mathbb{N}$ for the trip time of this intersection.
\begin{algorithm}
 \caption{Coordination Pseudocode for CAV $i$$\in\mathcal{N}_r(t)$}
 \begin{flushleft}
        \textbf{Input:} $\mathcal{T}_i=\left[\underline{t}_i^{f}, \overline{t}_i^f\right]$, planned trajectory of CAVs in $\mathcal{N}_r(t)$\\
        \textbf{Output:} {$t_i^f$}
\end{flushleft}

\begin{algorithmic}[1]
\State{$t_{i}^f \gets \underline{t}_i^f$}
\While{$t_{i}^f \leq \overline{t}_i^f $}
    \State{$k \gets$ CAV physically located in front of CAV $i$ }
    \While{$p_{k}(t)-p_i(t) < \delta_i(t)$}
    \Comment{Constraint \eqref{eqn:rearend1}}
        \State{$t_{i}^f \gets t_{i}^f+dt$}
    \EndWhile
    \State{$j \gets$ CAV physically located behind of CAV $i$ }
    \While{$p_{i}(t)-p_j(t) < \delta_j(t)$}
    \Comment{Constraint \eqref{eqn:rearend2}}
    \State{$t_{i}^f \gets t_{i}^f+dt$}
    \EndWhile
    \State{$\mathcal{L} \gets$ list of CAVs with the potential of lateral collision}
    \For{$k \in~$ $\mathcal{L}$}
    \State{$c$ $\gets$ conflict node between $i$ and $k$}
    \State{$a$ $\gets$ initialize with a negative number}
    \State{$b$ $\gets$ initialize with a positive number}
    \While{$\min \{a,b\}>0$}
    \Comment{Constraint \eqref{eq:lateralMinSafety}}
    \State{Compute $t_{i}^c$ based on current $t_i^f$}
    \State{$a$ $\gets$ $\max\{ \delta_i(t) + p_i(t) - p_i^c\},~\forall t\in[t_i^0, t_k^c]$}
    \State{$b$ $\gets$ $\max \{ \delta_k(t) + p_k(t) - p_k^c \}   ,~\forall t\in[t_k^0, t_i^c]$}
    \State{$t_{i}^f \gets t_{i}^f+dt$}
    \EndWhile
\EndFor
\State{\texttt{Return} $t_{i}^f$}
\EndWhile
\State{\texttt{Return} $-1$}
\end{algorithmic} \label{Alg:Coordination}
\end{algorithm}

\section{Simulation} \label{sec:simulation}

To evaluate the effectiveness of the proposed framework, we compare a baseline scenario using a road network with $70$ nodes and $198$ edges, which includes $20$ intersections and $4$ stations.
For the baseline scenario, we assume that each CAV updates its route at the beginning of each trip, i.e., pick-up trip, service trip, and return trip,  based on the traffic condition at that moment.
After finding the route, we use coordination of CAVs along the given routes to compute the actual travel time of the baseline scenario.

\begin{figure}[t]
    \centering
    \includegraphics[width=1.05\linewidth]{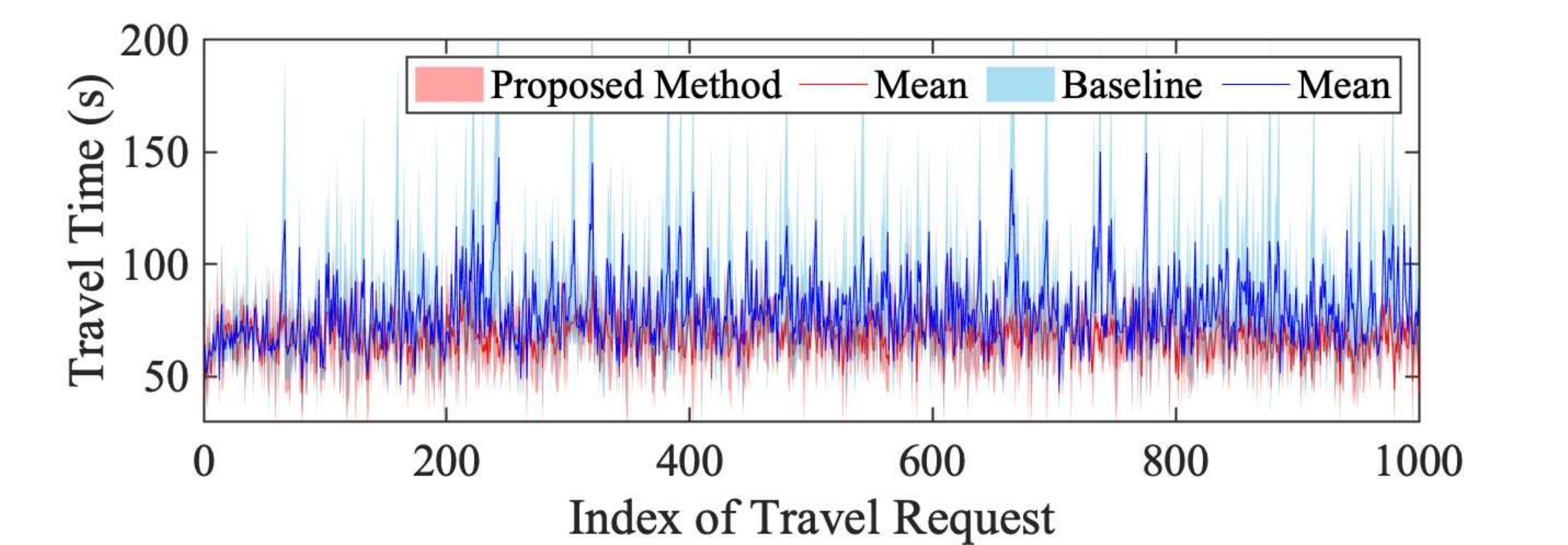}
    \caption{Travel time of each travel request for the baseline scenario (blue) and the proposed method (red).}
    \label{fig:travel}
\end{figure}

We conducted five simulations to compare the results.
For each simulation, we randomly generated $1000$ different travel requests.
Figure \ref{fig:travel} illustrates the travel time of each travel request.
The shaded area presents the travel time of all five simulations, and the solid line is the average time of those results.
This figure shows that our framework is not affected by the number of CAVs significantly.
On the other hand, the travel time of the baseline scenario has some fluctuation since the CAVs can only receive traffic information at certain moments.
The total travel time of our method was $18.7$ hours, while that of the baseline scenario was $22.1$ hours.
This result implies that a combined routing and coordination framework could save a significant amount of time, and probably energy, than deriving the optimal routing separated and then considering the coordination of CAVs.

\begin{figure}
    \centering
    \includegraphics[width=1.05\linewidth]{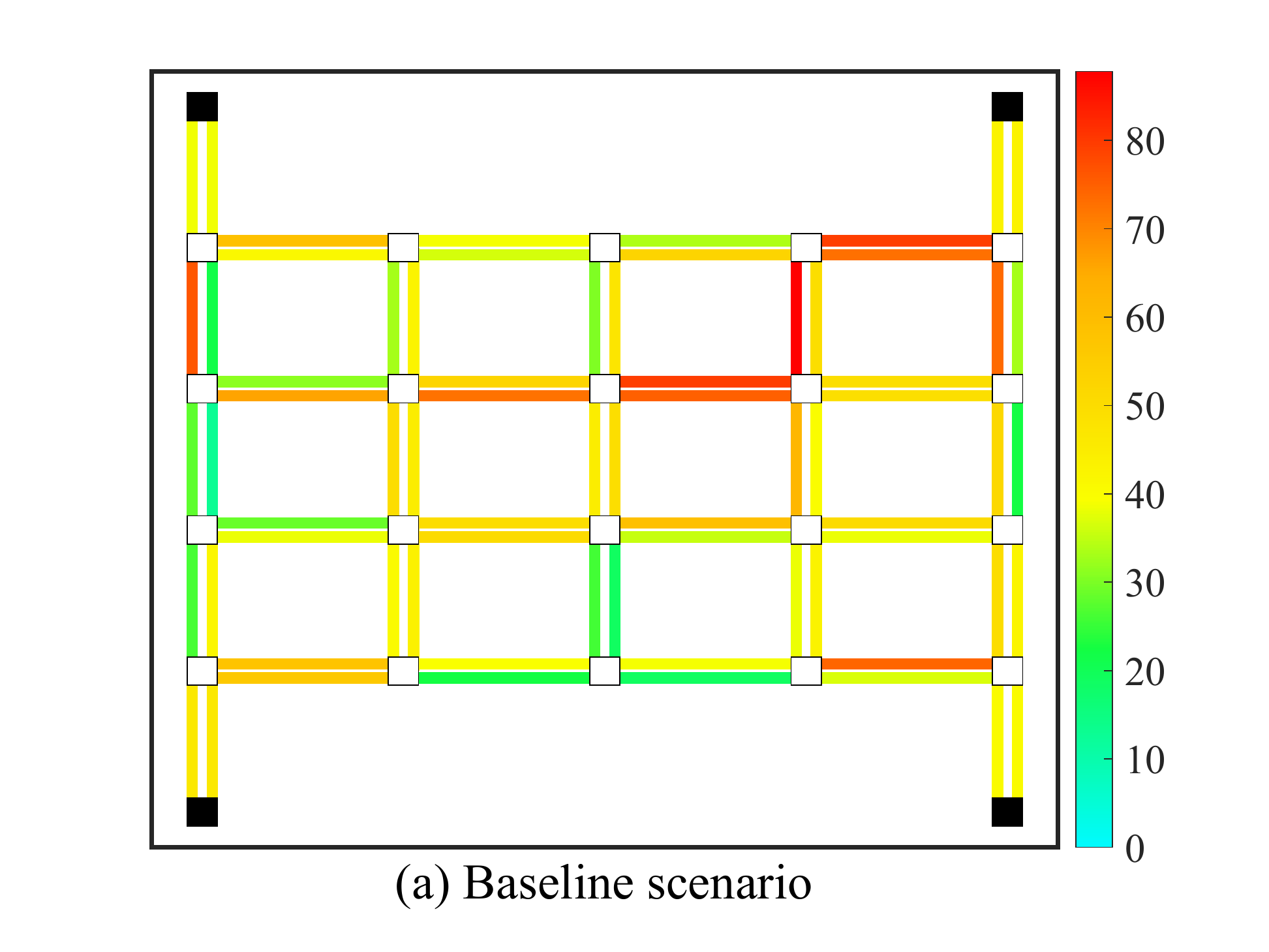}\label{subfig:baseline}\\
    \includegraphics[width=1.05\linewidth]{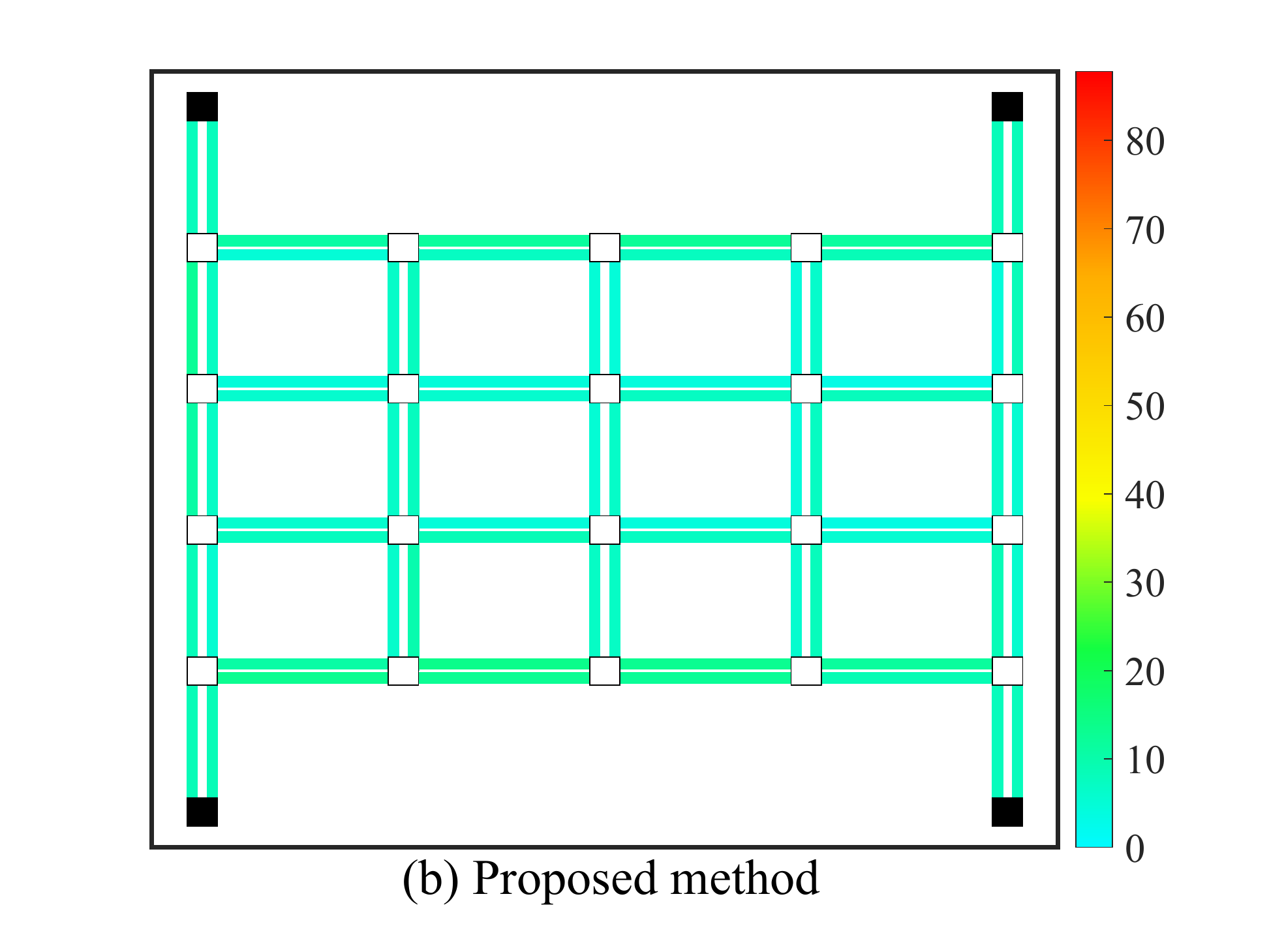}\label{subfig:proposed}
    
    \caption{Average road usage of CAVs with (a) baseline scenario and (b) proposed method.}
    \label{fig:roadusage}
\end{figure}

Figure \ref{fig:roadusage} visualizes the average road usage of CAVs for a selected simulation, where white and black squares are intersections and stations, respectively. 
It can be seen that our framework results in utilizing all the roads in the network to keep the low congestion level, while the baseline scenario has some focused road usage.

\section{Concluding Remarks} \label{sec:conclusion}
In this letter, we proposed an optimal routing framework combined with coordinating CAVs at signal-free intersections.
We formulated a shortest-time routing problem along with an optimization problem at each intersection which yields the energy-optimal trajectory of each CAV crossing the intersection. 
We demonstrated the effectiveness of our framework through simulation and showed significant advantages of combing the routing problem with the coordination of CAVs problem.

This is the beginning of a long journey, and as such, there are several interesting directions for future research. For example, future research should consider different traffic scenarios, e.g., merging at roadways and roundabouts, cruising in congested traffic, and lane-merging or passing maneuvers.
In addition, in this letter, we only considered the ideal case with a 100\% penetration rate of CAVs and signal-free intersections. Thus, future research should include more realistic traffic conditions such as multiple lanes, signalized intersections, and mixed traffic with human-driven vehicles.

\bibliographystyle{IEEEtran}
\bibliography{Bang, IDS, SharedMobilityRef,TAC_references}

\end{document}